\documentclass[twocolumn,aps,pra,showpacs]{revtex4}
\begin{document}
\title{Two qubits of a $W$ state violate
Bell's inequality beyond Cirel'son's bound}
\author{Ad\'{a}n Cabello}
\email{adan@us.es}
\affiliation{Departamento de F\'{\i}sica Aplicada II,
Universidad de Sevilla, 41012 Sevilla, Spain}
\date{\today}


\begin{abstract}
It is shown that the correlations between two qubits selected from
a trio prepared in a $W$ state violate the
Clauser-Horne-Shimony-Holt inequality more than the correlations
between two qubits in any quantum state. Such a violation beyond
Cirel'son's bound is smaller than the one achieved by two qubits
selected from a trio in a Greenberger-Horne-Zeilinger state [A.
Cabello, Phys. Rev. Lett. {\bf 88}, 060403 (2002)]. However, it
has the advantage that all local observers can know from their own
measurements whether or not their qubits belongs to the selected
pair.
\end{abstract}


\pacs{03.65.Ud,
03.65.Ta}
\maketitle


\section{Introduction}
\label{sec:I}


The Bell inequality \cite{Bell64} proposed by Clauser, Horne,
Shimony, and Holt (CHSH) \cite{CHSH69}, points out that in any
local-realistic theory, that is, in any theory in which the local
variables of a particle determine the results of local experiments
on this particle, the absolute value of a combination of four
correlations is bound by 2,
\begin{equation}
|C(A,B) - m C(A,b)-n C(a,B) - m n C(a,b)| \le 2.
\label{CHSH}
\end{equation}
In inequality~(\ref{CHSH}), $A$ and $a$ are two observables taking
values $-1$ or $1$ on particle $i$, and $B$ and $b$ are two
observables taking values $-1$ or $1$ on a distant particle $j$;
$m$ and $n$ can be either $-1$ or 1.

The CHSH inequality~(\ref{CHSH}) is violated for certain quantum
states and certain choices of the observables $A$, $a$, $B$, and
$b$ \cite{CHSH69}. Therefore, the conclusion is that no
local-realistic theory can reproduce the predictions of quantum
mechanics \cite{Bell64}.

Later on, Cirel'son \cite{Cirelson80} showed that, according to
quantum mechanics, for any two-qubit system {\em prepared in a
quantum state}, the absolute value of the combination of
correlations appearing in the CHSH inequality~(\ref{CHSH}) is
bound by $2 \sqrt{2}$ ({\em Cirel'son's bound}). This bound is
also the maximum violation predicted by quantum mechanics for the
two-qubit singlet state (or any other two-qubit Bell state)
\cite{CHSH69}. Indeed, this is the violation of Bell's inequality
traditionally tested in real experiments involving systems of two
qubits prepared in a quantum state \cite{ADR82,
SA88,RT90,KMWZSS95,WJSHZ98,RKVSIMW01}.

However, as shown in Ref.~\cite{Cabello02a}, according to quantum
mechanics the CHSH inequality can be violated beyond Cirel'son's
bound. The reason is the following. Bell's inequalities are
derived assuming local realism, without any mention to quantum
mechanics. Therefore, when searching for violations of a Bell's
inequality, one is not restricted to studying correlations between
ensembles of systems {\em prepared in a quantum state}; instead,
one can study any ensemble of systems, irrespective of whether
such an ensemble is meaningful in quantum mechanics or not (i.e.,
irrespective of whether it can be described by a quantum state or
not). For instance, one can consider trios of qubits prepared in a
certain quantum state and then assume local realism to select a
pair of qubits in each trio, and calculate, using quantum
mechanics, the correlations between these two qubits. The whole
procedure makes sense and can be translated into real experiments
as long as one can obtain the required correlations and
probabilities for the two selected qubits from the data obtained
in a real experiment with three qubits prepared in a quantum
state.

In Ref.~\cite{Cabello02a}, a violation of the CHSH
inequality~(\ref{CHSH}) beyond Cirel'son's bound is presented for
certain subensembles of two qubits of an ensemble of trios
prepared in a Greenberger-Horne-Zeilinger (GHZ) state
\cite{GHZ89}. In this paper, we shall investigate whether a
violation beyond Cirel'son's bound could be found for pairs of
qubits selected from trios prepared in a $W$ state \cite{DVC00}.

The structure of the paper is as follows. In Sec.~II, we will find
that, for a certain choice of observables, the $W$ state violates
the CHSH inequality~(\ref{CHSH}) beyond Cirel'son's bound. The
observables used in Sec.~II do not provide the maximum achievable
violation using a $W$ state. In Sec.~III, we explain the reason
behind this choice of observables. In addition, the violation in
Sec.~II is smaller that the one obtained in Ref.~\cite{Cabello02a}
using a GHZ state. However, in Sec.~IV, we will see that there are
some reasons that make the violation provided by the $W$ state
more interesting than that provided by the GHZ state. Finally, in
Sec.~V, we discuss how to obtain the required probabilities for
the two selected qubits from the data obtained in a real
experiment with three qubits.


\section{The $W$ state violates the CHSH inequality
beyond Cirel'son's bound}
\label{sec:II}


Let us consider three distant qubits 1, 2, 3, prepared in the $W$ state
\begin{equation}
\left| {W} \right\rangle = {1 \over {\sqrt{3}}}
\left( \left| +-- \right\rangle + \left| -+- \right\rangle +
\left| --+ \right\rangle \right),
\label{Wstate}
\end{equation}
where $\sigma_z \left| \pm \right\rangle = \pm \left| \pm
\right\rangle$. For each three qubits prepared in the $W$
state~(\ref{Wstate}), we are going to concentrate our attention on
{\em two} of them, namely, {\em those two in which, if we had
measured $\sigma_z$, we would have obtained the result $-1$}.
These two qubits will be called $i$ and $j$ hereafter, while the
corresponding third qubit (the one in which, if we had measured
$\sigma_z$, we would have found the result $1$) will be called
$k$. In quantum mechanics, the result of measuring $\sigma_z$ is
not predefined and therefore this prescription for choosing pairs
is meaningless. However, the prescription makes sense in a
local-realistic theory.

For reasons that will be explained in Sec.~III, we are interested
in the correlations when we choose $A=Z_i$, $a=X_i$, $B=Z_j$, and
$b=X_j$, where $Z_q$ and $X_q$ are the spin of qubit $q$ along the
$z$ and $x$ directions, respectively. In addition, the particular
CHSH inequality~(\ref{CHSH}) we are interested in is the one in
which $m=n=x_k$, where $x_k$ is one of the possible results, $-1$
or $1$ (although we do not know which one), of measuring $X_k$.
With this choice we obtain the following CHSH inequality:
\begin{equation}
|C(Z_i,Z_j) - x_k C(Z_i,X_j)
-x_k C(X_i,Z_j) - C(X_i,X_j)| \le 2,
\label{CHSH23}
\end{equation}
which holds for any local-realistic theory, regardless of the
particular value, either $-1$ or $1$, of $x_k$.

The next step is to use quantum mechanics to calculate the four
correlations appearing in inequality~(\ref{CHSH23}) for the
subensemble of two qubits $i$ and $j$ taken from three qubits
prepared in the $W$ state~(\ref{Wstate}).

For the subensemble of two qubits $i$ and $j$ defined above,
\begin{equation}
C(Z_i,Z_j)= 1,
\label{ZZ}
\end{equation}
because, for the $W$ state~(\ref{Wstate}),
\begin{eqnarray}
P_{Z_1 Z_2 Z_3} \left( {1,-1,-1} \right)
& = & {1 \over 3},
\label{5a} \\
P_{Z_1 Z_2 Z_3} \left( {-1,1,-1} \right)
& = & {1 \over 3},
\label{5b} \\
P_{Z_1 Z_2 Z_3} \left( {-1,-1,1} \right)
& = & {1 \over 3},
\label{5c}
\end{eqnarray}
where $P_{Z_1 Z_2 Z_3} \left( {1,-1,-1} \right)$ means the
probability of qubit $1$ giving the result $1$, and qubits $2$ and
$3$ giving the result $-1$ when measuring $\sigma_{z}$ on all
three qubits.

By the definition of qubits $i$ and $j$,
\begin{equation}
C(Z_i,X_j)= -x_k,
\label{ZX}
\end{equation}
because, for the $W$ state~(\ref{Wstate}),
\begin{eqnarray}
\!\!\!\!\!\!\!\!P_{Z_1 X_2 X_3} \left( {-1,1,-1} \right) +
P_{Z_1 X_2 X_3} \left( {-1,-1,1} \right)& = & 0,
\label{W1} \\
\!\!\!\!\!\!\!\!P_{X_1 Z_2 X_3} \left( {1,-1,-1} \right) +
P_{X_1 Z_2 X_3} \left( {-1,-1,1} \right)& = & 0,
\label{W2} \\
\!\!\!\!\!\!\!\!P_{X_1 X_2 Z_3} \left( {1,-1,-1} \right) +
P_{X_1 X_2 Z_3} \left( {-1,1,-1} \right)& = & 0.
\label{W3}
\end{eqnarray}
Analogously, using Eqs.~(\ref{W1})--(\ref{W3}),
\begin{equation}
C(X_i,Z_j)= -x_k.
\label{XZ}
\end{equation}
Finally,
qubit $k$ is the one in which, if we had measured
$\sigma_z$, we would have found the result $1$. The other two are qubits $i$ and $j$.
For the $W$ state~(\ref{Wstate}),
\begin{eqnarray}
P \left( X_2=X_3 | Z_1=1 \right) & = &  P \left( X_2=-X_3 | Z_1=1
\right),\\
P \left( X_1=X_3 | Z_2=1 \right) & = &  P \left( X_1=-X_3 | Z_2=1
\right),\\
P \left( X_1=X_2 | Z_3=1 \right) & = &  P \left( X_1=-X_2 | Z_3=1
\right),
\end{eqnarray}
where $P \left( X_2=X_3 | Z_1=1 \right)$
is the conditional probability of $X_2$ and $X_3$ having the same
result, given that the result of $Z_1$ is $1$.
Therefore,
irrespective of whether $i$ and $j$ are qubits $2$ and $3$, or $1$
and $3$, or $1$ and $2$, we conclude that
\begin{equation}
C(X_i,X_j)= 0.
\label{XX}
\end{equation}

Correlations~(\ref{ZZ}), (\ref{ZX}), (\ref{XZ}), and (\ref{XX})
violate the CHSH inequality~(\ref{CHSH23}). The violation (3 vs 2)
goes beyond Cirel'son's bound ($2 \sqrt{2}$).


\section{Why $X$ and $Z$?}
\label{sec:III}


A particular type of local-realistic theories are those in which
the only local experiments whose results are assumed to be
predetermined are those which satisfy the criterion for ``elements
of reality'' proposed by Einstein, Podolsky, and Rosen (EPR): {\em
``If, without in any way disturbing a system, we can predict with
certainty (i.e., with probability equal to unity) the value of a
physical quantity, then there exists an element of physical
reality corresponding to this physical quantity''} \cite{EPR35}.

As can be easily checked, the violation reported in Sec.~II is not
the maximal violation of the CHSH inequality~(\ref{CHSH23}) for
two qubits in the $W$ state~(\ref{Wstate}). For instance,
considering local spin observables on plane $x$-$z$ and assuming
$A=B$ and $a=b$, we find a maximum violation of 3.046 [by choosing
$A=\cos(0.628) \sigma_{x} - \sin(0.628) \sigma_{z}$ and
$a=\cos(1.154) \sigma_{x} + \sin(1.154) \sigma_{z}$]. Why then
have we chosen $A=Z_i$, $a=X_i$, $B=Z_j$, and $b=X_j$? The reason
is that these observables are not only local observables but, for
the $W$ state~(\ref{Wstate}), they also satisfy EPR's criterion of
elements of reality.

From Eqs.~(\ref{5a})--(\ref{5c}), it can be immediately seen that
$z_1$, $z_2$, and $z_3$ are elements of reality, since any of them
can be predicted with certainty from spacelike separated
measurements of $\sigma_z$ on the other two qubits. In addition,
from Eqs.~(\ref{W1})--(\ref{W3}), it can easily be seen that, if
$z_i=-1$ then, with certainty, $x_j=x_k$. Therefore, if $z_i=-1$,
then by measuring $x_j$ ($x_k$) one can predict $x_k$ ($x_j$) with
certainty. Therefore, if $z_i=-1$, then $x_j$ and $x_k$ are
elements of reality. If $z_i=1$ then, using
Eqs.~(\ref{5a})--(\ref{5c}), it can immediately be seen that
$z_j=-1$. Therefore, following the previous reasoning, $x_i$ and
$x_k$ are elements of reality (although $x_i$ could have ceased to
be an element of reality after measuring $\sigma_z$ on particle
$i$). In conclusion, for trios of qubits in the $W$
state~(\ref{Wstate}), $z_1$, $z_2$, $z_3$, $x_1$, $x_2$, and $x_3$
are EPR elements of reality and thus, according to EPR, they
should have predefined values $-1$ or $1$ before any measurement.

The violation of the CHSH inequality~(\ref{CHSH23}) presented in
Sec.~II is thus not only a proof of the impossibility of local
hidden variables, but also proves a more powerful result: the
apparently mild condition proposed by EPR is inconsistent with
quantum mechanics.


\section{Why $W$?}
\label{sec:IV}


As was shown in Ref.~\cite{Cabello02a}, two qubits belonging to
three-qubit system in a GHZ state can provide a higher violation
(4 vs 2, instead of 3 vs 2) of the CHSH inequality~(\ref{CHSH23}),
even using observables that satisfy EPR's criterion of elements of
reality. Why then use a $W$ state?

One reason is because a test of the violation of Bell's
inequalities beyond Cirel'son's bound could be achieved in
practice in the near future. Sources of $W$ states based on
parametric down-converted photons are now available for real
experiments \cite{Harald02} and some new proposals to prepare $W$
states via cavity quantum electrodynamics have recently been
presented \cite{GZ02}.

Another reason is because this violation beyond Cirel'son's bound
is, in one sense, surprising. The $W$ state is the genuine
three-qubit entangled state whose entanglement has the highest
robustness against the loss of one qubit \cite{DVC00}. In
particular, from a single copy of the reduced density matrix for
any two qubits belonging to a three-qubit $W$ state, one can
always obtain by means of a filtering measurement a state that is
arbitrarily close to a Bell state. Therefore, one might think that
any two qubits belonging to a $W$ state will not lead to a higher
violation of the CHSH inequality~(\ref{CHSH23}) than that for two
qubits in a Bell state, and thus it is of interest to realize that
this is not the case.

There is, however, another subtler reason for preferring the $W$
state instead of the GHZ state for a test of violation of Bell's
inequalities beyond Cirel'son's bound. Any test of this kind
requires a prescription for selecting a pair of qubits from each
trio prepared in a quantum state. Such a prescription assumes
local realism. In the violation of the CHSH
inequality~(\ref{CHSH23}) presented in Sec.~II, this prescription
is simple: qubits $i$ and $j$ are those two in which, if we had
measured $\sigma_z$, we would have obtained the result $-1$.
However, in the violation of the CHSH inequality~(\ref{CHSH23})
using a GHZ state described in Ref.~\cite{Cabello02a}, the
prescription is not so simple: there, qubits $i$ and $j$ are
either those two in which, if we had measured $\sigma_z$, we would
have obtained the result $-1$, or any two, if we had obtained the
result $1$ for all three qubits if we had measured $\sigma_z$.
This means that, for the $W$ state, any local observer could know
whether or not his qubit belonged to the selected pair just by
measuring $\sigma_z$; while for the GHZ state, the fact that a
qubit belongs or not to the selected pair cannot be decided with
certainty from a measurement on that qubit, but requires knowledge
of the results of measurements on the other two qubits. From the
perspective of local realism, for the $W$ state, one of the
elements of reality carried by each qubit determines whether or
not it belongs to the selected pair; while for the GHZ state, this
information is not local since it is distributed among distant
elements of reality.


\section{Experimental CH inequality}
\label{sec:V}


The result in Sec.~II opens the possibility of using sources of
three-qubit $W$ states \cite{Harald02,GZ02} to experimentally test
the CHSH inequality. The main advantage of an experiment like this
(or that proposed in Ref. \cite{Cabello02a}) is that it will admit
a direct comparison with the dozens of previous experiments with
two qubits \cite{ADR82, SA88,RT90,KMWZSS95,WJSHZ98,RKVSIMW01} and
thus goes beyond any previous experiments to test local realism
using sources of three qubits \cite{BPDWZ99,PBDWZ00} inspired by
proofs of Bell's theorem without inequalities \cite{GHZ89} or by
Bell's inequalities for three qubits \cite{Mermin90c,RS91}.

However, in any real experiment using three qubits, the
experimental data consist on the number of simultaneous detections
by three detectors $N_{ABC} \left( {a,b,c} \right)$ for various
observables $A$, $B$, and $C$. This number is assumed to be
proportional to the corresponding joint probability, $P_{ABC}
\left( {a,b,c} \right)$. Therefore, in order to make
inequality~(\ref{CHSH23}) useful for real experiments, it would be
convenient to translate it into the language of joint
probabilities.

Taking into account that
\begin{eqnarray}
P_{Z_i Z_j} \left( {-1,-1} \right) & = & {1 \over 4} [1-C(Z_i)-C(Z_j)
\nonumber \\ & & +C(Z_i,Z_j)],
\label{pro1} \\
P_{Z_i X_j} \left( {-1,-x_k} \right) & = & {1 \over 4} [1-C(Z_i)-x_k C(X_j)
\nonumber \\ & & +x_k C(Z_i,X_j)],
\label{pro2} \\
P_{X_i Z_j} \left( {-x_k,-1} \right) & = & {1 \over 4} [1-x_k C(X_i)-C(Z_j)
\nonumber \\ & & +x_k C(X_i,Z_j)],
\label{pro3} \\
P_{X_i X_j} \left( {x_k,x_k} \right) & = & {1 \over 4} [1+x_k C(X_i)+x_k C(X_j)
\nonumber \\ & & +x_k^2 C(X_i,X_j)],
\label{pro4}
\end{eqnarray}
where $C(Z_i)$ is the mean of the results of measuring $\sigma_z$
on qubit $i$, and assuming physical locality [i.e. assuming that
$C(Z_i)$ is independent of whether $\sigma_z$ or $\sigma_x$ is
measured on qubit $j$, that is, assuming that the value of
$C(Z_i)$ is the same in Eqs.~(\ref{pro1}) and (\ref{pro2}), etc.],
the CHSH inequality~(\ref{CHSH23}) between correlations can be
transformed into a Clauser-Horne (CH) inequality \cite{CH74}
between joint probabilities,
\begin{eqnarray}
-1\!\!& \le &\!\!P_{Z_i Z_j} \left( {-1,-1} \right) - P_{Z_i X_j} \left(
{-1,-x_k} \right) \nonumber \\
& &\!\!-P_{X_i Z_j} \left( {-x_k,-1}
\right)-
 P_{X_i X_j} \left( {x_k,x_k} \right) \le 0.
\label{CH23}
\end{eqnarray}
As can be easily checked, the bounds $l$ of the CHSH
inequality~(\ref{CHSH23}) are transformed into the bounds
$(l-2)/4$ of the corresponding CH inequality~(\ref{CH23}).
Therefore, the local-realistic bound in the CH
inequality~(\ref{CH23}) is 0 and Cirel'son's bound is
$(\sqrt{2}-1)/2 \approx 0.207$.

For qubits $i$ and $j$ of a system in the $W$ state~(\ref{Wstate}),
\begin{eqnarray}
P_{Z_i Z_j} \left( {-1,-1} \right) & = & 1,
\label{pzz} \\
P_{Z_i X_j} \left( {-1,-x_k} \right) & = & 0,
\label{pzx} \\
P_{X_i Z_j} \left({-x_k,-1} \right) & = & 0,
\label{pxz} \\
P_{X_i X_j} \left( {x_k,x_k} \right) & = & {3 \over 4}.
\label{pxx}
\end{eqnarray}
Therefore, probabilities~(\ref{pzz})--(\ref{pxx}) violate the CH
inequality~(\ref{CH23}). Such a violation (0.25 vs 0) is beyond
the corresponding Cirel'son's bound (0.207).

On the other hand, since we do not know which ones are qubits $i$
and $j$, we cannot obtain the four joint
probabilities~(\ref{pzz})--(\ref{pxx}) just by performing
measurements on two qubits. Therefore, we must show how the joint
probabilities of qubits $i$ and $j$ are related to the
probabilities of the three qubits.

As can easily be seen from the definition of qubits $i$ and $j$,
\begin{eqnarray}
P_{Z_i Z_j} \left( {-1,-1} \right) & = & P_{Z_1 Z_2 Z_3} \left(
{1,-1,-1} \right) \nonumber \\
& & +P_{Z_1 Z_2 Z_3} \left({-1,1,-1} \right) \nonumber \\
& & +P_{Z_1 Z_2 Z_3} \left({-1,-1,1} \right) \nonumber \\
& & +P_{Z_1 Z_2 Z_3} \left({-1,-1,-1} \right).
\label{PZZ}
\end{eqnarray}
Therefore, in order to experimentally obtain $P_{Z_i Z_j} \left(
{-1,-1} \right)$, we must measure the four probabilities in the
right-hand side of Eq.~(\ref{PZZ}). In the $W$
state~(\ref{Wstate}), the first three probabilities in the
right-hand side of Eq.~(\ref{PZZ}) are expected to be $1/3$ and
the fourth is expected to be zero.

On the other hand, $P_{Z_i X_j} \left( {-1,-x_k} \right)$ and
$P_{X_i Z_j} \left( {-x_k,-1} \right)$ are both less than or equal
to
\begin{eqnarray}
& &\!\!\!\!P_{Z_1 X_2 X_3} \left( {-1,1,-1} \right)+P_{Z_1 X_2 X_3}
\left( {-1,-1,1} \right) \nonumber \\
& &\!\!\!\!+P_{X_1 Z_2 X_3} \left(
{1,-1,-1} \right)+P_{X_1 Z_2 X_3} \left( {-1,-1,1} \right)
\nonumber \\
& &\!\!\!\!+P_{X_1 X_2 Z_3} \left( {1,-1,-1} \right)+P_{X_1 X_2 Z_3} \left(
{-1,1,-1} \right).
\label{sum}
\end{eqnarray}
Therefore, in order to experimentally obtain $P_{Z_i X_j} \left(
{-1,-x_k} \right)$ and $P_{X_i Z_j} \left( {-x_k,-1} \right)$, we
must measure (using three different setups) all six probabilities
in sum~(\ref{sum}). In the $W$ state~(\ref{Wstate}), each of these
six probabilities is expected to be zero.

Finally,
\begin{eqnarray}
P_{X_i X_j} \left( {x_k,x_k} \right)
& = & P_{X_1 X_2 X_3} \left({1,1,1} \right) \nonumber \\
& & +P_{X_1 X_2 X_3} \left({-1,-1,-1} \right).
\label{PXX}
\end{eqnarray}
Therefore, in order to experimentally obtain $P_{X_i X_j} \left(
{x_k,x_k} \right)$, we must measure the two probabilities in the
right-hand side of Eq.~(\ref{PXX}). In the $W$
state~(\ref{Wstate}), each of them is expected to be $3/8$.


\section{Conclusions}


Two qubits selected from a trio prepared in a $W$ state violate
the CHSH inequality, or the corresponding CH inequality, more than
two qubits prepared in {\em any} quantum state. Such violations
beyond Cirel'son's bound are smaller than those achieved by two
qubits selected from a trio in a GHZ state \cite{Cabello02a}.
However, for the $W$ state the argument is simpler, since all
local observers can know from their own measurements whether or
not their qubits belong to the selected pair.

The importance of these arguments relies on the fact that they
suggest how to use sources of three-qubit quantum entangled states
to experimentally reveal violations of the familiar two-qubit Bell
inequalities beyond those obtained using sources of two-qubit
quantum states.


\begin{acknowledgments}
This work was prompted by a question made by H.~Weinfurter during
the Conference {\em Quantum Information:
Quantum Entanglement} (Sant Feliu de Gu\'{\i}xols, Spain, 2002).
I thank D.~Collins for pointing out a mistake in a
previous version, C.~Serra for
comments, and the Spanish Ministerio de Ciencia y Tecnolog\'{\i}a,
Grant No.~BFM2001-3943, and the Junta de Andaluc\'{\i}a, Grant
No.~FQM-239, for support.
\end{acknowledgments}


\end{document}